\documentclass[a4paper,11pt]{article}
\pdfoutput=1

\usepackage{jinstpub}

\usepackage{array,booktabs}
\usepackage{float}
\usepackage{subcaption}

\title{\boldmath Radiation hardness of a p-channel notch CCD developed for the X-ray CCD camera onboard the XRISM satellite}

\author[a,1]{Y. Kanemaru \note{Corresponding author.}}
\author[a]{J. Sato}
\author[a]{K. Mori}
\author[b]{H. Nakajima}
\author[a]{Y. Nishioka}
\author[a]{A. Takeda}
\author[c,d]{K. Hayashida}
\author[c,d]{H. Matsumoto}
\author[c]{J. Iwagaki}
\author[c]{K. Okazaki}
\author[c]{K. Asakura}
\author[c]{T. Yoneyama}
\author[e]{H. Uchida}
\author[e]{H. Okon}
\author[e]{T. Tanaka}
\author[e]{T. G. Tsuru}
\author[f]{H. Tomida}
\author[f]{T. Shimoi}
\author[g]{T. Kohmura}
\author[g]{K. Hagino}
\author[h]{H. Murakami}
\author[i]{S. B. Kobayashi}
\author[a]{M. Yamauchi}
\author[a]{I. Hatsukade}
\author[j]{M. Nobukawa}
\author[k]{K. K. Nobukawa}
\author[l]{J. S. Hiraga}
\author[m]{H. Uchiyama}
\author[n]{K. Yamaoka}
\author[f]{M. Ozaki}
\author[f]{T. Dotani}
\author[c]{H. Tsunemi}
\author[o]{T. Hamano}

\affiliation[a]{Faculty of Engineering, University of Miyazaki, 1-1 Gakuen Kibanadai Nishi, Miyazaki, Miyazaki 889-2192, Japan}
\affiliation[b]{Faculty of Science and Engineering, Kanto Gakuin University, 1-50-1 Mutsuura Higashi, Kanazawa-ku, Yokohama, Kanagawa 236-8501, Japan}
\affiliation[c]{Department of Earth and Space Science, Graduate School of Science, Osaka University, 1-1 Machikaneyama-cho, Toyonaka, Osaka 560-0043, Japan}
\affiliation[d]{Project Research Center for Fundamental Sciences, Graduate School of Science, Osaka University, 1-1 Machikaneyama-cho, Toyonaka, Osaka 560-0043, Japan}
\affiliation[e]{Department of Physics, Kyoto University, Kitashirakawa Oiwake-cho, Sakyo-ku, Kyoto, Kyoto 606-8502, Japan}
\affiliation[f]{Japan Aerospace Exploration Agency, Institute of Space and Astronautical Science, 3-1-1 Yoshino-dai, Chuo-ku, Sagamihara, Kanagawa 252-5210, Japan}
\affiliation[g]{Department of Physics, Faculty of Science and Technology, Tokyo University of Science, 2641 Yamazaki, Noda, Chiba 270-8510, Japan}
\affiliation[h]{Department of Information Science, Faculty of Liberal Arts, Tohoku Gakuin University, 2-1-1 Tenjinzawa, Izumi-ku, Sendai, Miyagi 981-3193, Japan}
\affiliation[i]{Department of Physics, Faculty of Science, Tokyo University of Science, Kagurazaka, Shinjuku-ku, Tokyo 162-0815, Japan}
\affiliation[j]{Faculty of Education, Nara University of Education, Takabatake-cho, Nara, Nara 630-8528, Japan}
\affiliation[k]{Department of Mathematical and Physical Sciences, Graduate School of Science, Nara Women's University, Kitauoyanishi-machi, Nara, Nara 630-8506, Japan}
\affiliation[l]{Department of Physics, Faculty of Science and Technology, Kwansei Gakuin University, 2-2 Gakuen, Sanda, Hyogo 669-1337, Japan}
\affiliation[m]{Faculty of Education, Shizuoka University, 836 Ohya, Suruga-ku, Shizuoka 422-8529, Japan}
\affiliation[n]{Department of Physics, Graduate School of Science, Nagoya University, Furo-cho, Chikusa-ku, Nagoya, Aichi 464-8602, Japan}
\affiliation[o]{Department of Accelerator and Medical Physics, National Institute of Radiological Sciences, 4-9-1 Anagawa, Inage-ku, Chiba 263-8555, Japan}

\emailAdd{kanemaru@astro.miyazaki-u.ac.jp}

\abstract{We report the radiation hardness of a p-channel CCD developed
for the X-ray CCD camera onboard the XRISM satellite. This CCD has basically the same
characteristics as the one used in the previous Hitomi satellite, but newly employs
a notch structure of potential for signal charges by increasing the implant
concentration in the channel . The new device was exposed up to approximately $7.9
\times 10^{10}$~protons~cm$^{-2}$ at 100~MeV. The charge transfer inefficiency
was estimated as a function of proton fluence with an ${}^{55}$Fe source. A
device without the notch structure was also examined for comparison.  The result
shows that the notch device has a significantly higher radiation hardness
than those without the notch structure including the device adopted for Hitomi.
This proves that the new CCD is radiation tolerant for space applications with a
sufficient margin.}


\proceeding{PIXEL 2018 Internatinal workshop
	\\
  December 10-14, 2018 
	\\
  Activity Center of Academia Sinica, Taipei, Taiwan 
}

\begin{document}
\maketitle
\flushbottom

\section{Introduction}
The X-Ray Imaging and Spectroscopy Mission (XRISM), recently renamed from
XARM, is the seventh Japanese X-ray astronomical satellite planned to be launched
in the early 2020's \cite{Tashiro2018}. XRISM will carry two
identical X-ray mirror assemblies. One of the focal plane detectors is an X-ray
microcalorimeter array, which will provide unprecedented high-resolution X-ray
spectroscopy with a relatively narrow field of view (FOV) of $3' \times 3'$ \cite{Ishisaki2018}.
The other is an X-ray charge-coupled device (CCD) camera, which has moderate energy
resolution with a large FOV of $38' \times 38'$ \cite{hayashida2018}. These two instruments play
complementary roles to each other and will open up a new view of the X-ray universe.

The XRISM CCD has basically the same characteristics as the one used in the previous
Hitomi satellite \cite{Takahashi2018}, 
a p-channel back-illuminated device with a full-depletion layer 
with a thickness of 200~$\mu$m. 
As is the case with Hitomi, XRISM will fly in the
low earth orbit with an altitude of 575~km and an inclination angle of
$31^{\circ}$. Devices in this orbit are exposed to a large number of cosmic rays,
dominated by geomagnetically trapped protons in the South Atlantic Anomaly, and the
average dose rate of protons is estimated to be $260~\mathrm{rad~year}^{-1}$ in
the case of the Hiromi CCD \cite{Mori2013}. 
The non-ionizing energy loss of cosmic-ray protons results
in bulk damage in silicon. It increases the charge transfer inefficiency (CTI)
defined as a fraction of charge loss per one-pixel transfer and degrades
the spectroscopic performance of X-ray CCDs in space.

In the case of Hitomi, in order to mitigate the radiation damage effects, we cooled
the CCD temperature down to $-110^\circ$C and employed the charge injection (CI)
technique, which reduces signal packet loss by filling traps with regularly spaced
injected charges \cite{Nakajima2018,tanaka2018}. For further
improvement, we newly introduced a ``notch'' structure to the XRISM CCD. The notch
structure is a narrow implant in the CCD channel confining a charge packet to a
fraction of the pixel volume in an additional potential well and has been known to
reduce the CTI \cite{bebek2002,tsunemi2004}.

In this paper, we report the results of radiation damage experiments for studying
the radiation hardness of our new notch device, especially paying attention to the
application to the XRISM satellite.

\section{Experiment}
\begin{figure}[h]
	\begin{minipage}{0.5\linewidth}
		\centering \captionof{table}{Specifications and operation parameters of the CCDs under test}
		\label{table:ccdspec}
		\begin{tabular}{cc}
			\toprule
			\midrule
			architecture		& frame-transfer \\
			channel type		& p-channel \\
			clock phase		& 2 \\
			pixel size 				& $ 24 \times 24~\mu \mathrm{m}^2$ \\
			pixel format      & 320(H) $\times$ 256(V) \\
			imaging area size & 7.7 $\times$ 6.1~mm$^{2}$\\
			\hline
			binning 				& 2 $\times$ 2 \\
			frame cycle		& 4~s \\
			operating temperature 	& $-110~{}^{\circ}\mathrm{C} $\\
			\bottomrule
		\end{tabular}
	\end{minipage}
	\begin{minipage}{0.5\linewidth}
		\centering
		\includegraphics[width=1\linewidth]{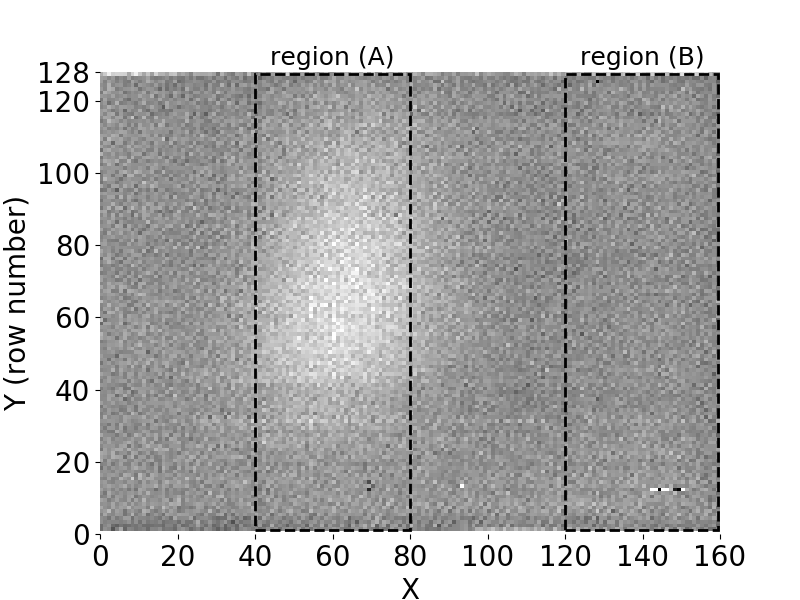}
		\caption{Dark current distribution of the notchless CCD after the irradiation. 
			The lighter the color, the higher the dark current value.}
	    \label{fig:mini08-15darkcrop}
	\end{minipage}
\end{figure}
Table \ref{table:ccdspec} shows the specifications and operation parameters of the
CCDs under test. The devices are the same as the flight model except for their
smaller pixel format \cite{hayashida2018}.  
Since the on-chip $2 \times 2$ binning is applied, the frame format obtained is
effectively a quarter of the pixel format.
In order to evaluate the effect of the notch
structure, we fabricated two CCDs. One device has a notch structure, and the other
does not. We hereafter call them ``notch CCD'' and ``notchless CCD''.

The radiation damage experiments were performed at HIMAC, which is a synchrotron
facility for heavy ion therapy at the National Institute of Radiological Sciences in
Japan. 
The beamline used in the experiment was PH1, which can provide 
a proton pencil beam with transverse profile approximated by Gaussian-shape
with a standard deviation of $\sim$ 1~mm, much smaller than the CCD size of $\sim$ 7~mm.
The beam of 100~MeV protons was
directly incident on the devices under atmospheric pressure and at room
temperature. We repeated the same experiment for the notch and notchless CCDs, and the
numbers of incident protons were $5.64 \times 10^9$ and $3.38 \times 10^9$,
respectively. After the irradiation, CTI was measured with an ${}^{55}$Fe source at
$-110~{}^{\circ}\mathrm{C}$ in our laboratory. Figure \ref{fig:mini08-15darkcrop}
represents the dark current distribution of the notchless CCD after the experiment.
The notch CCD also showed a similar profile to the distribution. 
It is clear that pixels with higher dark current are localized
due to the concentration of the proton beam around the center of the imaging
area. In the following analysis, we focus on region (A) in $40 \le
\mathrm{X} \le 80$ and (B) in $120 \le \mathrm{X} \le 160$ to represent 
severely and scarcely damaged areas, respectively (Figure~\ref{fig:mini08-15darkcrop}).

\section{Analysis \& Result}
\begin{figure}[h]
	\begin{center}
		\begin{tabular}{ccc}
			\begin{tabular}{c}
				Notch \\
				CCD
			\end{tabular} &
			\begin{minipage}{0.4\linewidth}
				\centering
				Region (A)
				\includegraphics[width=1\linewidth,clip]{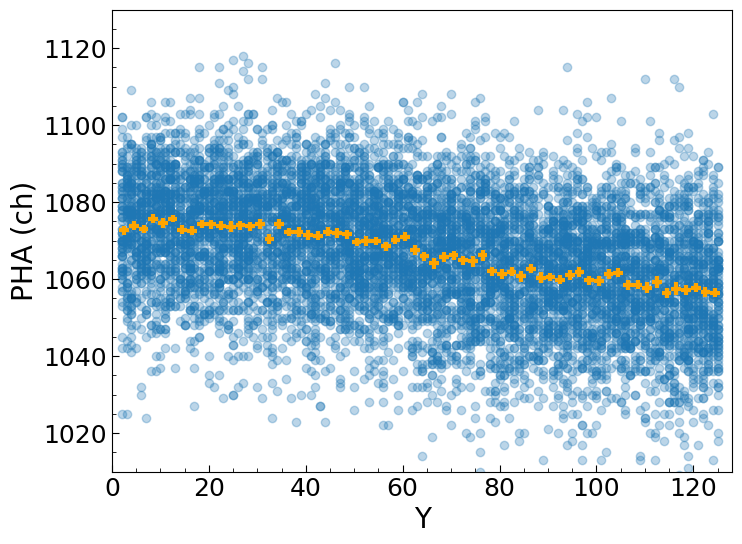}
			\end{minipage} &
			\begin{minipage}{0.4\linewidth}
				\centering
				Region (B)
				\includegraphics[width=1\linewidth,clip]{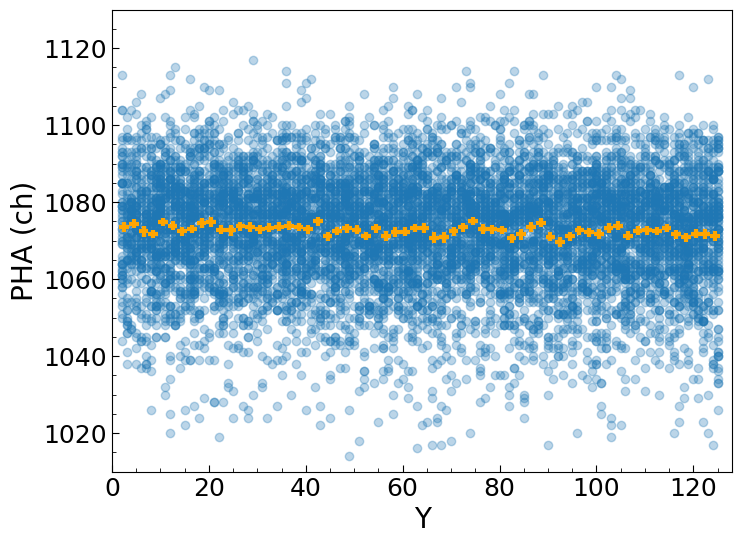}
			\end{minipage}\\
			\begin{tabular}{c}
				Notchless \\
				CCD
			\end{tabular} &
			\begin{minipage}{0.4\linewidth}
				\begin{center}
					\includegraphics[width=1\linewidth,clip]{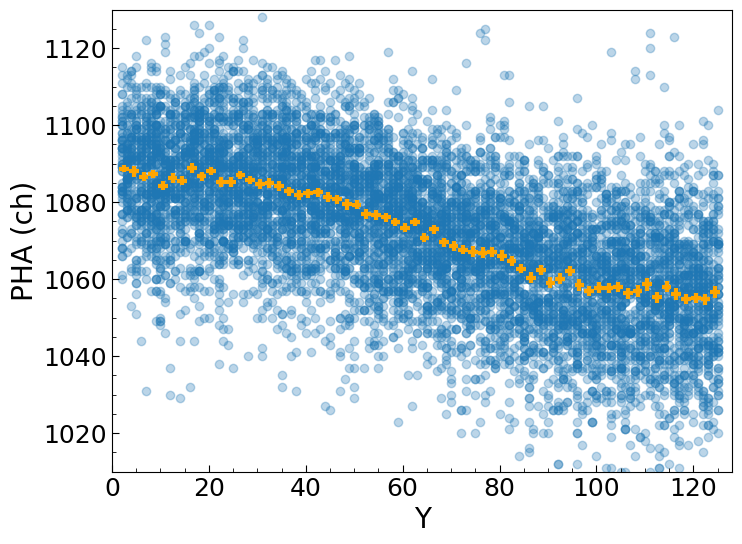}
				\end{center}
			\end{minipage} &
			\begin{minipage}{0.4\linewidth}
				\begin{center}
					\includegraphics[width=1\linewidth,clip]{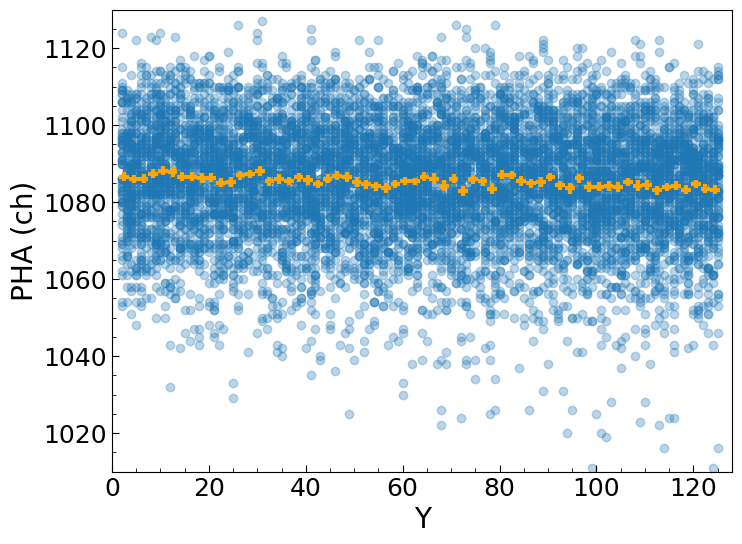}
				\end{center}
			\end{minipage}
		\end{tabular}
	\caption{ Pulse heights of X-ray event produced by the Mn-K$\alpha$ line from
	 an ${}^{55}$Fe source as a function of the row number of Y. Blue dots show
	 each event and the yellow crosses denote the mean of the pulse height every
	 2 rows and the standard deviation of the mean. Upper and lower panels show
	 those of the notch and notchless CCDs, and left and right panels show those
	 in region (A) and region (B), respectively.  } \label{fig:stackingplots}
	\end{center}
\end{figure}

Figure \ref{fig:stackingplots} shows the pulse heights of X-ray events produced by
the Mn-K$\alpha$ line from an ${}^{55}$Fe source as a function of the row number of
Y.  The Y value
corresponds to half the number of transfers because of the $2 \times 2$
binning. The single pixel events in which signal charges are confined in one pixel
are used.  In region (B), where the proton fluence was almost zero, pulse
heights barely reduce with increment in the number of transfers. On the other hand,
the events in region (A), where the beam was incident, apparently and
non-linearly lost charges as the number of transfers increases. Comparing the notch
and notchless CCD results (comparing upper and lower panels), the
pulse height reduction of the notch CCD is smaller in spite of the larger total
number of the incident protons to the notch CCD. It qualitatively indicates that
the CTI degradation of the notch CCD is mitigated by employing the notch structure.

\subsection{Measurement of CTI}
Defining that $\mathit{CTI}_y$ is the value of the CTI in the charge transfer between the row
numbers of $y$ and $y-1$ in the 2$\times$2 binned format, CTI can be quantitatively
 evaluated by fitting the pulse
heights with the function of the row number as below:

\begin{align}
\mathit{PHA}(Y) &= \mathit{PHA}_0 \times (1 - \mathit{CTI}_1)^2 \times (1 - \mathit{CTI}_2)^2 \times \cdots \times (1 - \mathit{CTI}_Y)^2 \notag \\
&= \mathit{PHA}_0 \times \prod_{y=1}^Y (1 - \mathit{CTI}_y)^2 ~, \label{eq:pha}
\end{align}

\noindent where $Y$ is the
row number of a binned pixel at which X-ray is incident, $\mathit{PHA}_0$ is the
pulse height corresponding to the original charge produced by the Mn-K$\alpha$ line, and
$\mathit{PHA}(Y)$ is the pulse height observed at the binned pixel with the row number
$Y$.  If the $\mathit{CTI}_y$ were constant, the equation (\ref{eq:pha}) could be
simplified as $\mathit{PHA}(Y) = \mathit{PHA_0}\,(1 - \mathit{CTI}_y)^{2Y}$, which
well describes the experimental situation where the radiation damage was uniform
across the imaging area \cite{Mori2013}.  
Since the proton fluence differed in each row in this case, the simplified function
does not apply. Considering that the beam has a Gaussian-shape profile,
we assumed that $\mathit{CTI}_y$ is represented by the following Gaussian function:

\begin{equation}
\mathit{CTI}_y = c \exp \left\{ - \frac{(y-Y_0)^2}{2\sigma^2} \right\} + \mathit{CTI}_\mathrm{init} ~, \label{eq:cti}
\end{equation}
\noindent
where $c$ is the maximum CTI,
$Y_0$ is the center of the beam axis, $\sigma$ is the beam width, and
$\mathit{CTI}_\mathrm{init}$ is the initial value of the CTI measured before the
experiment.

\begin{figure}[h]
	\begin{subfigure}{0.5\linewidth}
		\centering
		\includegraphics[width=1\linewidth]{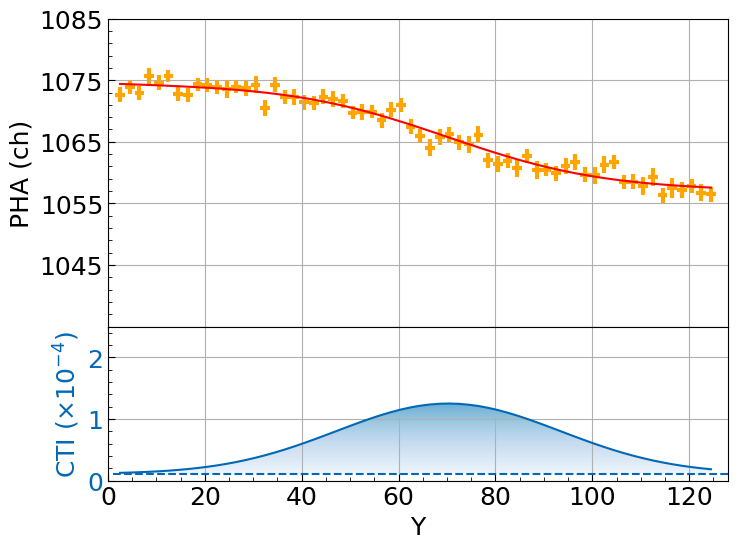}
		\caption*{Notch CCD}
	\end{subfigure}	
	\begin{subfigure}{0.5\linewidth}
		\centering
		\includegraphics[width=1\linewidth]{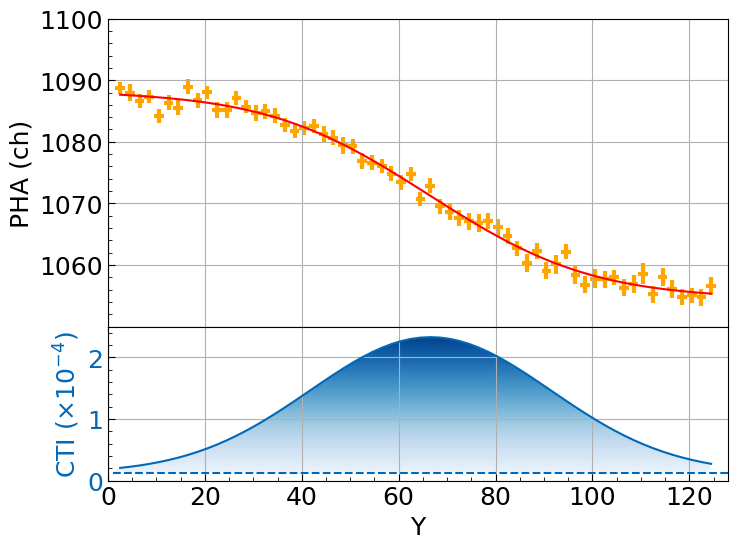}
		\caption*{Notchless CCD}
	\end{subfigure}	
	\caption{Vertical profile of pulse heights in region (A) (yellow cross)
 with the best fit function (red line) (upper panel) and 
 the $\mathit{CTI}_{y}$ (blue solid line) with the initial CTI value (blue dotted line) (lower panel)
 as a function of the row number of Y.
 The data of the pulse heights are
 the same as shown in figure~\ref{fig:stackingplots}. 
}
 \label{fig:CTIresults}
\end{figure}
Figure \ref{fig:CTIresults} shows the fit results. All datasets are well described by
the composite model consisting of the equations (\ref{eq:pha}) -- (\ref{eq:cti}), and
the parameters obtained are reasonable. For example, $Y_0$ in the notchless
CCD case was $66.6\pm 1.1$, which matches the peak of the dark current distribution
shown in figure \ref{fig:mini08-15darkcrop} and where the pulse height reduction is
the largest. The same applies to the notch CCD case.

\subsection{Estimation of the proton fluence in each row}
\begin{figure}[h]
	\begin{subfigure}{0.5\linewidth}
		\centering
		\includegraphics[width=1\linewidth]{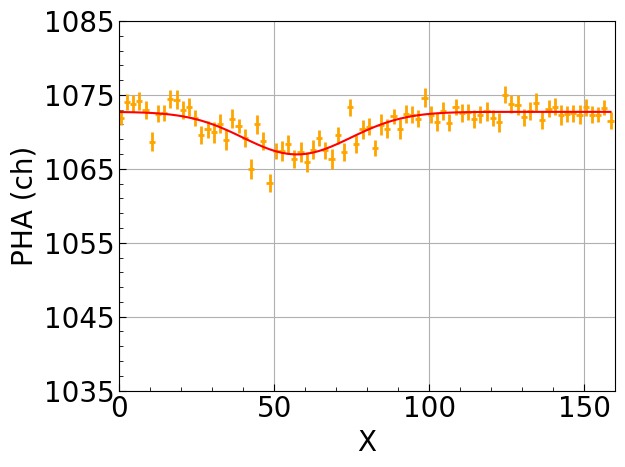}
		\caption*{Notch CCD}
	\end{subfigure}	
	\begin{subfigure}{0.5\linewidth}
		\centering
		\includegraphics[width=1\linewidth]{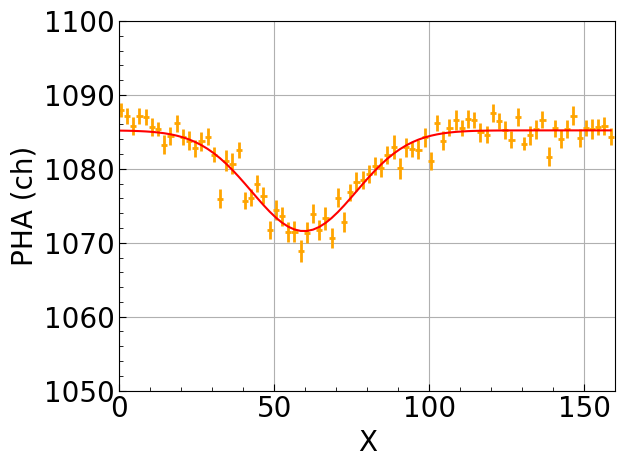}
		\caption*{Notchless CCD}
	\end{subfigure}	
	\caption{Horizontal profiles of pulse heights (yellow cross) with the best fit function (red line).}
	\label{fig:horizontalprofile}
\end{figure}

In order to quantify the relation between the CTI and the radiation damage at
each row, the proton fluence in each row also needs to be estimated.  Since the
total number of the incident protons to the imaging area was measured, the proton
fluence in the row of $y$ was estimated by integration of the beam distribution:
\begin{equation}
	{\Phi(y)} = n_p\int_{Y=y}^{Y=y+1}\int_{X=40}^{X=80} f(X, Y)\,dXdY ~,
\end{equation}
where $\Phi(y)$ is the proton fluence in the row of $y$, $n_p$ is the total protons
incident to the imaging area, and $f$ is a normalized beam distribution. We
approximated the beam distribution as a 2D Gaussian function.  The value of the
vertical width was taken from the CTI model fitting described above while that of
the horizontal width was estimated by fitting the horizontal profile of pulse
heights with the Gaussian function as shown in figure~\ref{fig:horizontalprofile}.
This horizontal profile was made from the single pixel events in the region of 40 $\le$ Y $\le$
80 where the damage was the most severe. The vertical and horizontal widths in the
notchless CCD case were $0.81 \pm 0.03$~mm and $1.19 \pm 0.08$~mm,
respectively. Similar values were obtained in the notch CCD case. These values were
consistent with those measured at the beam monitor in the upper stream of our
system \cite{torikoshi1999}, especially in terms of the ratio of the vertical and horizontal widths. In
our estimation, the notch and notchless CCDs were irradiated by up to ${\sim}~7.9
\times 10^{10}$~protons~cm$^{-2}$ and ${\sim}~4.5 \times 10^{10}$~protons~cm$^{-2}$
in the highest fluence area, respectively.

\subsection{Evolution of CTI as a function of the equivalent time in orbit}

\begin{figure}[h]
	\begin{subfigure}{0.5\linewidth}
		\centering
		\includegraphics[width=1\linewidth]{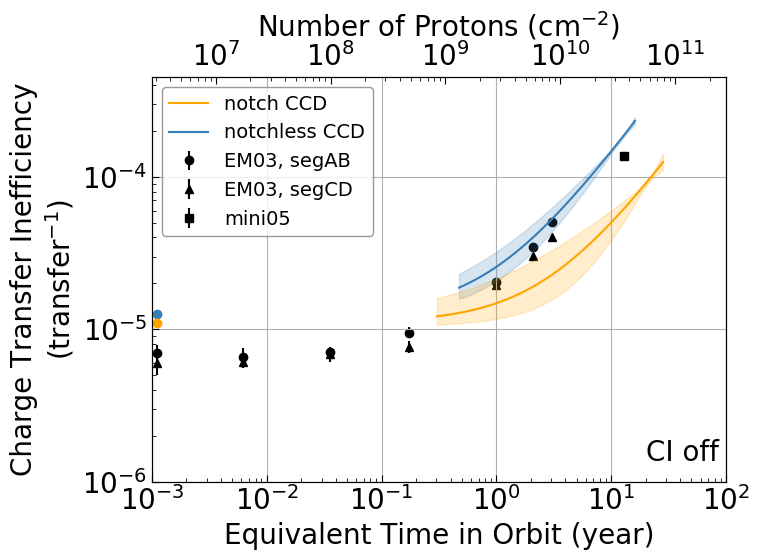}
		\caption*{w/o CI technique}
	\end{subfigure}	
	\begin{subfigure}{0.5\linewidth}
		\centering
		\includegraphics[width=1\linewidth]{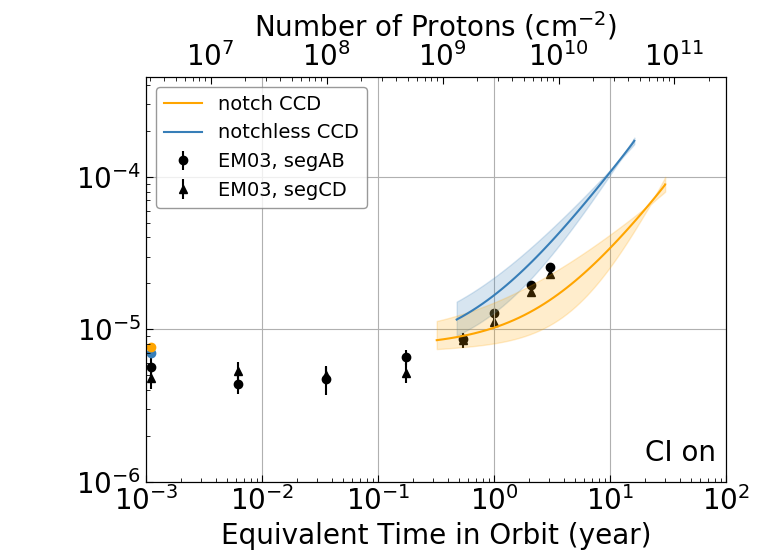}
		\caption*{w/ CI technique}
	\end{subfigure}	
	\caption{CTI as a function of equivalent time in orbit. The vertical axis is
	CTI and the horizontal axis on the bottom is the equivalent time in orbit, which is
	converted from the proton fluence on the top axis \cite{Mori2013}.  The blue
	and orange lines show the results of the notchless and notch CCDs,
	respectively. The blue and orange dots at $10^{-3}$~years indicate the initial CTI
	values before the experiments. The black dots show the results of our previous
	measurement for the Hitomi CCD \cite{Mori2013}.}  \label{CTIvsequivtime}
\end{figure}
Figure \ref{CTIvsequivtime} shows the CTI as a function of the equivalent time
in the low earth orbit where the XRISM satellite is planned to be injected. This
figure basically plots $\mathit{CTI}_{y}$ vs $\Phi(y)$ obtained above, and $\Phi(y)$
is converted to equivalent time in the XRISM orbit following Mori et al. (2013)
\cite{Mori2013}. It is clear that the introduction of the notch structure mitigates
the increase of CTI by a factor of 2--3 (comparison between blue and orange
lines). Comparison with our previous measurement of the notchless Hitomi CCD, 
which was performed with an $^{55}$Fe source at $-110^{\circ}$C, also
shows the effectiveness of the notch structure (comparison between black dots and
the orange line). Since we had already shown that even the notchless CCD adopted for
Hitomi was radiation tolerant enough for space use \cite{Mori2013}, these results
suggest that our new notch CCD has a sufficient margin of radiation tolerance for
the application to XRISM. We also confirmed that the CI technique effectively works
for both types of the CCDs (comparison between left and right figures).

\section{Discussion}

\begin{figure}[h]
	\begin{subfigure}{0.5\linewidth}
	\centering
	\includegraphics[width=1\linewidth]{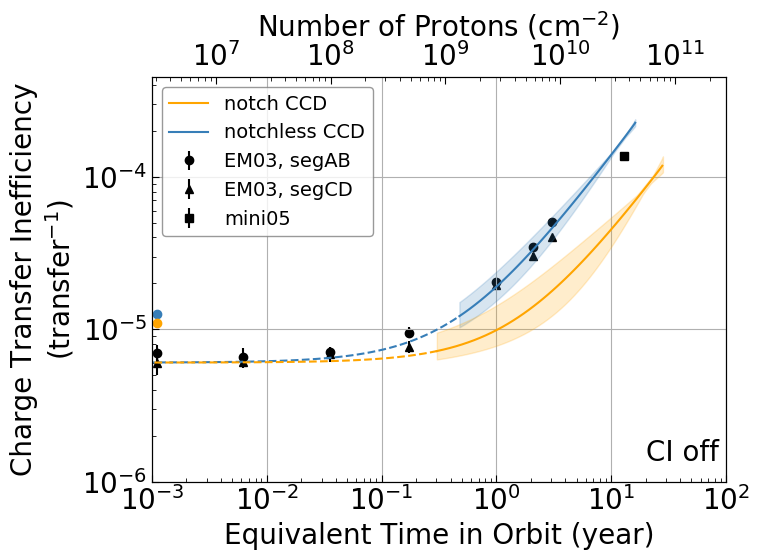}
		\caption*{w/o CI technique}
	\end{subfigure}	
	\begin{subfigure}{0.5\linewidth}
	\centering
	\includegraphics[width=1\linewidth]{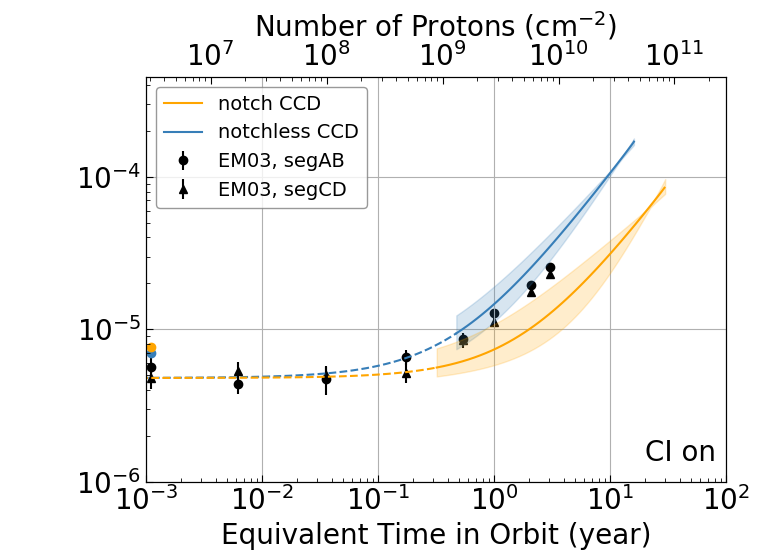}
	\caption*{w/ CI technique}
	\end{subfigure}	
	\caption{Same as figure \ref{CTIvsequivtime} but the initial CTI values
of the notch and notchless CCDs used in this experiment are hypothetically set to
the same value of the Hitomi CCD. The dotted lines are the eye guide to clarify the
 hypothetical situation.}
\label{CTIvsequivtimeCTIinit}
\end{figure}

We performed proton radiation damage experiments on our newly developed notch CCD and
previously developed notchless CCD, and verified the effectiveness of the notch structure in this
simple control experiment. The introduction of the notch structure improved radiation hardness of
our device by a factor of 2--3. Other experiments using different manufacturing p-channel CCDs
have also reported a similar degree of improvement from comparisons of their notch and notchless
devices \cite{bebek2002,marshall2004}. We note that experimental conditions,
such as proton beam energy and CCD working temperature, are different among experiments
including ours. Although the detailed manufacturing processes regarding the notch implant of each
device are not available, this might suggest that the width ratios between the notch implant and the channel are similar to each other.

The notchless CCD is basically the same as that adopted for Hitomi and thus it is expected
that their radiation hardness is comparable. However, in figure \ref{CTIvsequivtime}, the CTI degradation of the notchless CCD used in this experiment appears to be greater than or equal to that of the Hitomi CCD (comparison between the blue line and black dots). 
Figure \ref{CTIvsequivtimeCTIinit} is the same as figure \ref{CTIvsequivtime} but the
initial CTI values of the notch and notchless CCDs used in this experiment are hypothetically set
to the same value of the Hitomi CCD. Here, we only changed the initial CTI values and the rest of
the parameters in the equations (\ref{eq:pha})--(\ref{eq:cti}) are fixed to the best fit values. Although the hypothetical
notchless CCD curves are yet to correspond completely to the Hitomi CCD data, the initial CTI
value differences may be a part of the reasons for the difference between the notchless CCD and the Hitomi CCDs in figure \ref{CTIvsequivtime}.

\acknowledgments
%
This work was performed as a part of accelerator experiments of the Research Project at NIRS-HIMAC. We would like to express our thanks to HIMAC crews for their kind support throughout the experiments. This work was also supported by JSPS KAKENHI Grant Number 16H03983 (K.M.), and 15H03641, 18H01256 (H.N.). 
%


\bibliographystyle{JHEP}
\bibliography{bibtex/pixel2018_himac}
\end{document}